# Evaluation of a length-based method to estimate discard rate and the effect of sampling size

Erla Sturludottir[a], Gudjon Mar Sigurdsson[b] and Gunnar Stefansson[a]

[a]Science Institution, University of Iceland, Taeknigardur, Dunhagi 5, 107 Reykjavik, Iceland

[b]Marine and Freswater Research Institute, Skulagata 4, 101 Reykjavik, Iceland

## Abstract

The common fisheries policy aims at eliminating discarding which has been part of fisheries for centuries. It is important to monitor the compliance with the new regulations but estimating the discard rate is a challenging task, especially where the practise is illegal. The aim of this study was to review a length-based method that has been used to estimate the discard rate in Icelandic waters and explore the effects of different monitoring schemes. The length-based method estimates the minimum discard rate and the method of bootstrapping can be used to determine the uncertainty of the estimate. This study showed that the number of ships is the most important factor to consider in order to decrease the uncertainty.

# 1 Introduction

Discarding of fish has been part of fisheries for centuries (Alverson et al. 1994, Pálsson 2002) and is now generally viewed as a waste of valuable recourses. There can be economic reasons why fishers discard fish but they may also discard fish to comply with regulations (Kelleher 2005, Eliasen et al. 2014). It can be difficult to avoid unwanted species or sizes of fish and these are then caught as a by-catch when fishing for the targeted species.

Estimating discard rate in a fishery is a challenging task and is especially difficult where discarding is illegal. Most of the methods to estimate discarding are based on data obtained by observers who are allocated to ships where they monitor the amount discarded (Erzini et al. 2002, Borges et al. 2005, Rochet and Trenkel 2005). This is possible where discarding is not prohibited. The amount discarded can then be determined onboard the vessels. Usually these monitoring programs do not have 100% coverage of the fishery and therefore the amount discarded needs to be raised to the whole fishery. Various factors have been used as raising factors (Borges et al. 2005, Rochet and Trenkel 2005). The most common assumption when raising the discard to the total population is that it is proportional to catches or effort but this assumption may not always be valid (Rochet and Trenkel 2005) and Borges et al. (2005) recommend using fishing trips instead.

The common fisheries policy aims at eliminating discarding with the introduction of the landing obligation which will be gradually implemented from 2015 to 2019. This will make monitoring of compliance with the new regulations difficult as observers will not be able to determine the amount discarded onboard vessels anymore as the discarding practise has become illegal. Studies have shown that discarding still exist where a discard ban has been in place for many years (Pálsson et al. 2012). Fewer methods have been developed to estimate discarding that can be used where a discard ban is in place. Casey (1996) describes a model that can be used where only landings-at-age are known. This method uses knowledge about the selection of the fishing gear used in the fishery and assumptions regarding discarding practises. Pálsson (2003) developed a length-based method to estimate discard rate. This method uses length distributions acquired from observers which sample fish from the catch at sea and from landings ashore. It also depends on assumption of the discarding practise.

High variability in discard rate between fisheries, fishing trips and hauls, gears, seasons, years and areas makes it very difficult to estimate discarding. These lead to expensive monitoring programs with low precision (Rochet and Trenkel 2005). The uncertainty of the discard estimation can be high despite large sample size as the survey design includes intra-class correlation. Pennington and Volstad (1994) describe the intra-class correlation or the intra-cluster correlation as they call it. They state that fish sampled in a survey cannot be assumed to be sampled at random from the population because fish caught in close vicinity is more alike than fish caught at random from the general population. It is known that even low intra-class correlation greatly increases the uncertainty of an estimate compared to a complete random sampling. Studies have shown that it is more important to sample from large number of vessels or tows rather than take large sample of fish when estimating length distributions (Helle and Pennington, 2004; Singh et al., 2016).

The aim of this study was to review the method developed by Pálsson (2003) and to test what the effect of data used would have on the estimate and to compare different bootstrapping methods to determine the precision of the estimate. The aim of this study was also to investigate how sample sizes affect the precision of the discard rate estimate.

## 2 Methods

### 2.1 The historical data

Monitoring of discards in Icelandic waters began in 2001 when the Marine and Freshwater Research Institute (MFRI) and the Directorate of Fisheries in Iceland began to systematically sample data to estimate discards. Samples of fish were taken from landings when discarding had already occurred and at sea by observers before any discarding took place and the length of the sampled fish measured. Samples were taken from the most common fishing gears: bottom trawl, long line, gill net and Danish seine. The focus of the monitoring program has been on cod and haddock but other species have also been sampled but that has varied between years and fishing gears.

The sampling at sea was a three stage sampling scheme (Fig. 1): 1) samples are taken from randomly chosen ships, 2) within each ship samples are take from one or more tows, 3) each sample contains around 100 observations, i.e. fish chosen at random which are then length measured. The sampling ashore was similar except samples from individual tows could not be taken.

The third stage, the observations of fish length measurements form a cluster where fish caught together are more similar than fish drawn randomly from the population, this has been called the intra-cluster correlation (Pennington and Volstad, 1994). The same can be assumed about the tows, length distributions from tows within a ship are more alike than the length distributions between ships.

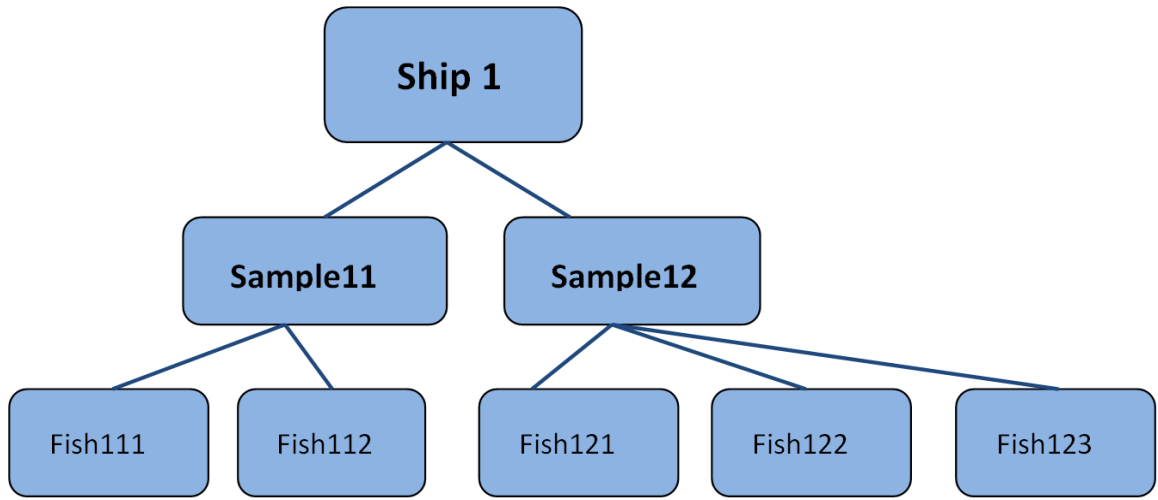

Figure 1. Three stage sampling scheme of the observations, i.e. length measurements of fish.

## 2.2 Estimation of discard rate

### 2.2.1 Data used in estimation

The historical data used for the discard rate estimation consist of samples from sea and ashore from different ships. Samples were often taken from a ship ashore and then later at sea but in some cases samples from a ship are only available from either sea or ashore. Also, the samples from a sea and ashore may not always be from the same locations. We compared three methods for choosing data: 1) use all available data to estimate discard rate, 2) use only samples from ships that have both samples from sea and ashore, and 3) only use samples from locations that have at least 90 observations from both at sea and ashore samples.

### 2.2.2 Length based estimation method

The discard rate was estimated from the historical data using a length-based method described in Pálsson (2003). The proportion discarded of each length group was estimated for cod and haddock caught by bottom trawl. Length ($L$) measurements taken at sea and ashore were used for the estimation. The individual weight ($W$) was assumed to follow the relationship

$$W = 0.01 \times L^3 \qquad (1)$$

The mean weight ($W_{mean}$) of all landed fish can then be calculated along with the total numbers landed ($N_{landed}$) using the total biomass ($B_{landed}$) of the landed catch:

$$N_{landed} = B_{landed}/W_{mean} \qquad (2)$$

The landed number in length class $l$ can then be estimated from the length distributions acquired from the samples. The total numbers caught at sea are not known as it is illegal to discard fish and are therefore not reported. However, the length distribution of the catch is available from samples taken at sea by observers. If fish is not discarded when it is larger than a certain length ($D_{max}$) the numbers caught can be calculated. First a raising factor ($k$) is calculated that can be used to raise the proportions to numbers of caught fish per length class. The $k$ is calculated as follows:

$$k = \frac{\sum_{l>D_{max}} N_{l,landed}}{\sum_{l>D_{max}} p_{l,catch}} \quad (3)$$

where $N_{l,landed}$ are numbers landed in length class $l$ and $p_{l,catch}$ is the proportion of each length class $l$ in the catch and $D_{max}$ represents the maximum length that is discarded. The total numbers caught ($N_{l,catch}$) in each length class can then be calculated as:

$$N_{l,catch} = k \times p_{l,catch} \quad (4)$$

The proportion discarded ($p_{l,discard}$) of each length class $l$ can then be calculated as:

$$p_{l,discard} = \frac{N_{l,catch} - N_{l,landed}}{N_{l,catch}} \quad (5)$$

To see how these equations come about, first write $N_{nondisc,landed} = \sum_{l>D_{max}} N_{l,landed}$ for the numbers landed in the non-discarded length groups and $N_{tot,landed} = \sum_l N_{l,landed}$ for the total number landed and if we knew the numbers caught we would write $N_{nondisc,catch} = \sum_{l>D_{max}} N_{l,catch}$ and $N_{tot,catch} = \sum_l N_{l,catch}$ for the corresponding values for caught fish, then the numbers caught in a length group are simply $N_{l,catch} = N_{tot,catch} \times p_{l,catch}$. We assume that no fish above a certain length ($D_{max}$) is discarded and therefore $N_{nondisc,landed} = N_{nondisc,catch}$ which is equivalent to $\sum_{l>D_{max}} N_{tot,landed} \times p_{l,landed} = \sum_{l>D_{max}} N_{tot,catch} \times p_{l,catch}$. We can derive the equation $N_{l,landed} = N_{tot,catch} \times p_{l,catch}$ for $l > D_{max}$. Summing this last equation over $l$ for stability, we see that $k$ in (4) is merely an estimate of the total numbers caught.

The proportion discarded of each length group ($\hat{p}_{l,discard}$) is assumed to follow a logistic curve:

$$\hat{p}_{l,discard} = \frac{1}{1 + \exp(-\hat{b}(L - \hat{D}_{50}))} \quad (6)$$

where $\hat{D}_{50}$ is the estimated length where the proportion discarded is 50% and $\hat{b}$ is the estimation of a parameter describing the steepness of the curve. The proportion discarded in each length group ($\hat{p}_{l,discard}$) was estimated for cod and haddock for bottom trawl for each year from 2001 to 2005.

### 2.2.3 Uncertainty using bootstrapping

The method described above does not give any uncertainty estimation. Uncertainty of an estimator can be estimated using the bootstrapping method (Efron and Tibshirani, 1993). The idea of this method is that you sample from the data with replacement instead from the population. If the data is a good representation of the population the estimate of the uncertainty will be realistic.

The data used for the discard estimation is hierarchal or multi-level data with three levels (ships, samples, observations). In this case the observations (length of fish) cannot be assumed to be independent as fish in the same sample or from the same ship are more alike than fish from different samples or ships. This is called the intra-class correlation or the intra-cluster correlation (Pennington and Volstad 1994, Ren et al. 2010).

There is no general method for resampling multi-level data (Ren et al. 2010). Resampling can be done at all three levels, two levels or at just one level, the highest level. Resampling at all levels would be as follows for our case: First the ships would be resampled with replacement, then the tows within each ship would be sampled with replacement, and finally the fish within each tow and ship would be sampled with replacement. Sampling from the highest level would be resampling the ships with replacement but the other levels without replacement. In this case each ship is taken as an observation. Ren et al. (2010) recommend resampling at one level, the highest level. In this study we will compare three methods: a) resampling at all three levels, b) resampling only at the highest level, and c) assuming all observations (fish measurements) as independent which has not been recommended. This will be done for all three data choosing methods (see Section 2.2.1). The 95% confidence interval was achieved by the percentile method (Efron and Tibshirani, 1993) of 5000 bootstrap samples.

## 2.3 *Effects of different monitoring schemes on the precision of the estimated discard rate*

Monte Carlo simulations were carried out to explore the effects of changing the number of ships, samples and observations on the precision of the discard rate estimate. The simulations were based on the method for bootstrapping hierarchical data (Ren et al. 2010, Singh et al. 2016) to take the intra-class correlation into account. Data sampled in 2002 in the discard monitoring project was used as the most extensive data was sampled that year. All available data was used (data usage method 1, see Section 2.2.1) to demonstrate the effects of changing sample size on the precision. Only observations from ships that had more than two samples (tows) and more than 100 fish per samples where used in the simulation. That resulted in 42 ships from sea and 23 ships from ashore for cod and 38 ships from sea and 26 ships from land for haddock. Therefore, the precision of the discard estimate was only considered for maximum of 40 ships at sea and 20 ships ashore, each with two samples with 100 observations.

The sampling from the dataset was done in such a way to preserve the intra-class correlation in the data. First, the ships were sampled with replacement, then samples within each ship were sampled with replacement and finally observations within each sample were sampled with replacement. This was done 5000 times for each combination of number of ships, samples and observations and the discard rate estimated. The precision was calculated as the range of the 95% confidence interval, i.e. the difference between the $0.025^{th}$ and the $0.975^{th}$ percentile of the 5000 estimates. The precision was estimated for 10, 20 and 40 ships at sea and 10 and 20 ships ashore where 1 or 2 samples were taken with various numbers of fish measurements (Table 1). All combinations of the at sea and ashore sampling were carried out, resulting in 72 monitoring schemes. This setup allows for comparisons with different numbers of ships and samples but same numbers of total observations. For example 10 ships, each with 2 samples with 25 observations has a total of 500 observation, the same as 20 ships with, each with 1 sample with 25 observations.

Table 1. Number of ships, samples and fish measurements at sea and ashore in the sampling procedure for estimating discard rate and precision.

| At sea | | | Ashore | | |
|---|---|---|---|---|---|
| N ship | N samples | N fish | N ship | N samples | N fish |
| 10 | 1 | 25 | 10 | 1 | 100 |
| 10 | 1 | 50 | 10 | 2 | 100 |
| 10 | 1 | 100 | 20 | 1 | 100 |
| 10 | 2 | 25 | 20 | 2 | 100 |
| 10 | 2 | 50 | | | |
| 10 | 2 | 100 | | | |
| 20 | 1 | 25 | | | |
| 20 | 1 | 50 | | | |
| 20 | 1 | 100 | | | |
| 20 | 2 | 25 | | | |
| 20 | 2 | 50 | | | |
| 20 | 2 | 100 | | | |
| 40 | 1 | 25 | | | |
| 40 | 1 | 50 | | | |
| 40 | 1 | 100 | | | |
| 40 | 2 | 25 | | | |
| 40 | 2 | 50 | | | |
| 40 | 2 | 100 | | | |

# 3 Results and discussion

## 3.1 *The discard rate estimation*

The estimated discard rate by numbers ranged from 0.06% in 2003 to 5.1% in 2005 for cod and from 1.0% in 2001 to 24.2% in 2003 for haddock (Fig. 2). There was some difference in the average discard rate depending on what data was used in the estimation. However, no method had always higher or lower values and the 95% confidence interval did show that there was not a significant difference of the three data usage methods as the confidence intervals did overlap for all methods for both cod and haddock (Fig. 2). The confidence interval was narrowest when all available data was used in the estimation for haddock discard rate but there was no consistent pattern for the range of the confidence interval for cod (Fig. 2). The discard rate of cod was not significantly different from 0 in the year 2003 for all the three data usage methods and from 2001 to 2003 where only data from same locations was used and for haddock the discard rate was not significantly different from 0 for all the three methods for the year 2001. The number of ships and samples were different between the years, which partly explains the difference in the range of the confidence interval. For example in 2001 only samples from seven ships ashore and samples from four ships at sea were used for the estimation when only using data from the same locations (Fig. 3). Note that the confidence interval is constrained at zero and therefore the confidence interval for the cod discard rate is narrower than for the haddock which has higher estimated discard rate than the cod.

It is not straight forward which data usage method is the most appropriate to use. Pálsson et al. (2012) only used data from locations which had more than 90 observations from samples taken ashore and at sea. In this case the length distributions acquired from samples ashore and at sea represent the same area. This does however not ensure that the samples have equal observations from each location within the area. For example, samples at sea may have higher frequency in locations with larger fish than samples taken ashore and will therefore have different length distributions. Using data from samples that come from ships that have both samples from ashore and at sea has been done when analyzing data for observer effect (Benoît and Allard, 2009). In this case the length distributions from the same ships can be compared. This does however not guarantee that the ashore and at sea samples come from the same locations. The samples from ashore and at sea should represent the same area but it may not be necessary that they come from same locations. To get more realistic length distributions it is possible to scale the samples at each location with total landings from each location. This can however not been done for the catch as the catch is not known at each location. The discard rate has been observed to be different between seasons but none of the methods take that into account.

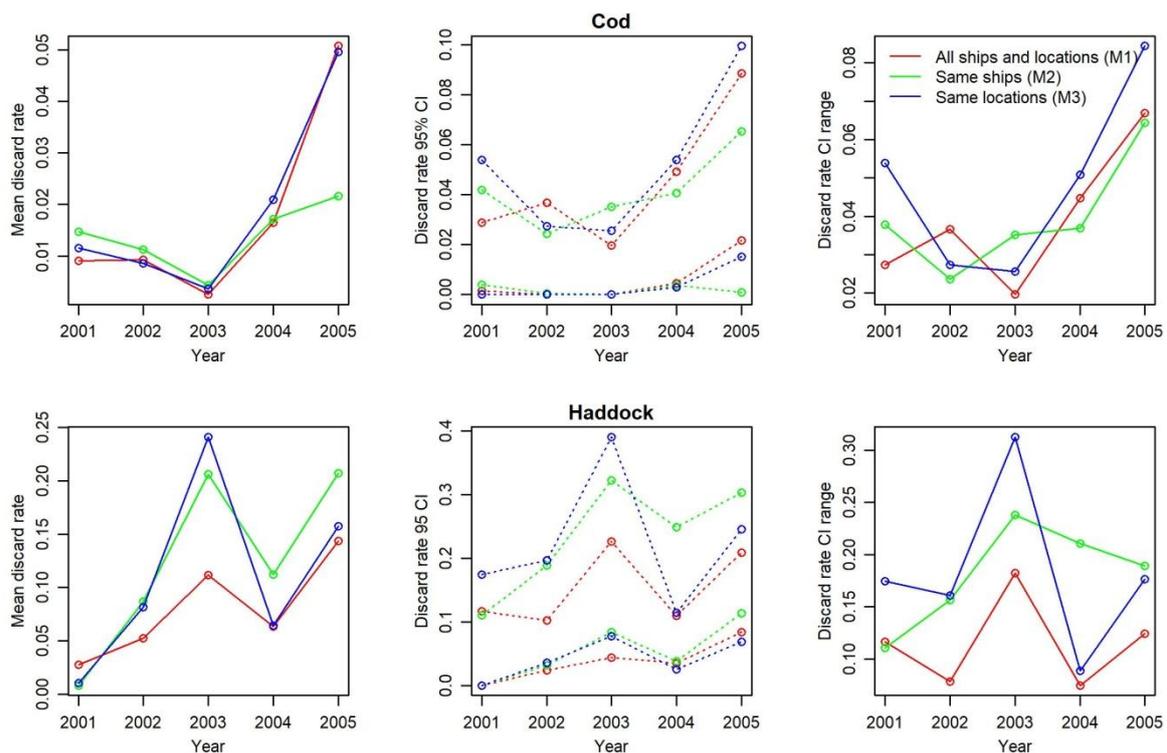

**Figure 2. Comparing the mean and 95% confidence interval (CI) of discard rate by numbers of the three data usage methods for cod and haddock: M1) All available data used, M2) Data from ships with both samples from sea and ashore used, and M3) only data from locations that have both sea and ashore samples. The 95% CI was achieved by resampling all three levels of data.**

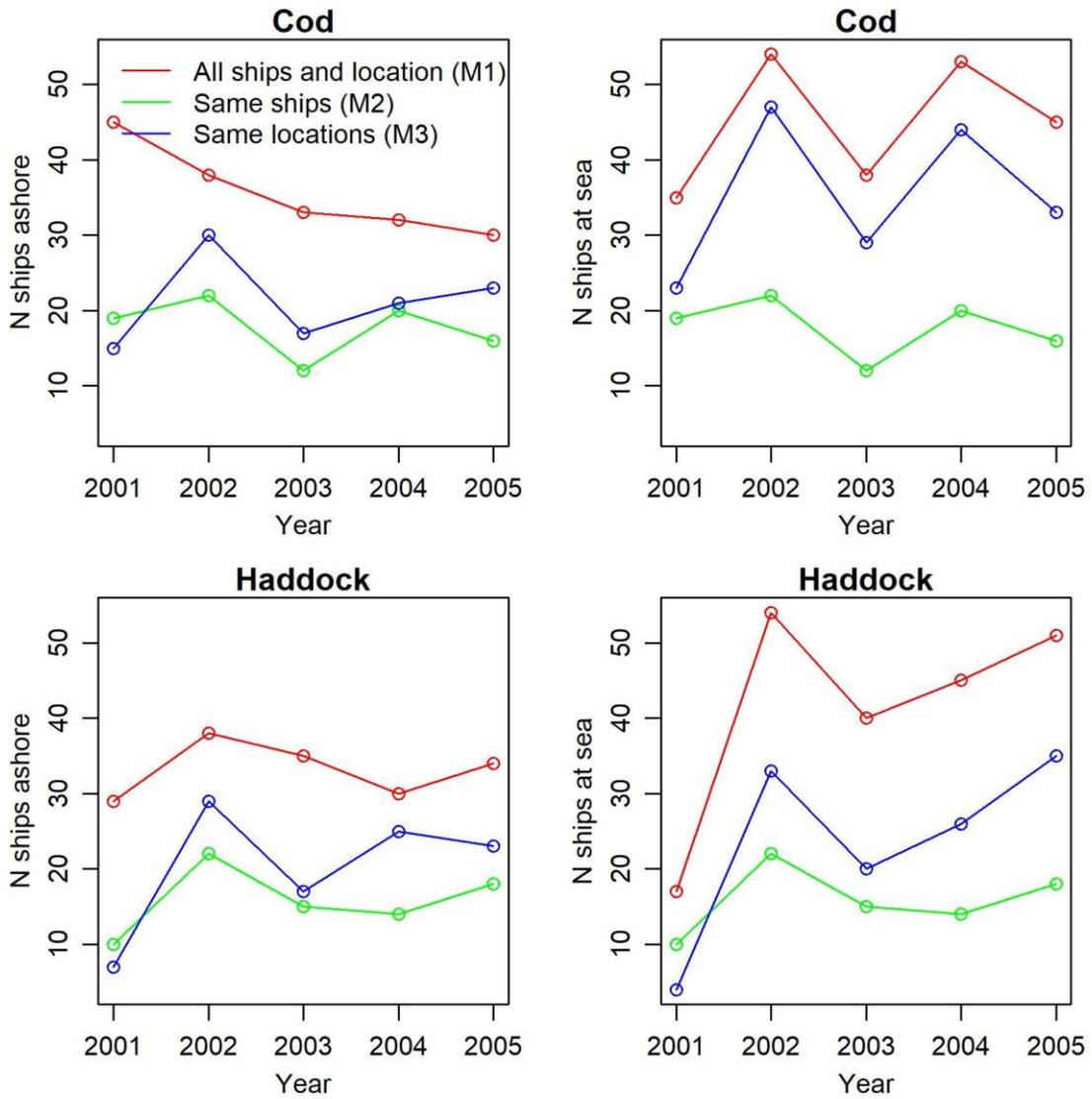

**Figure 3.** Number of ships at sea and ashore used for the estimation of discard rate for cod and haddock for the three methods: M1) All ships and locations, M2) same ships, and M3) same locations.

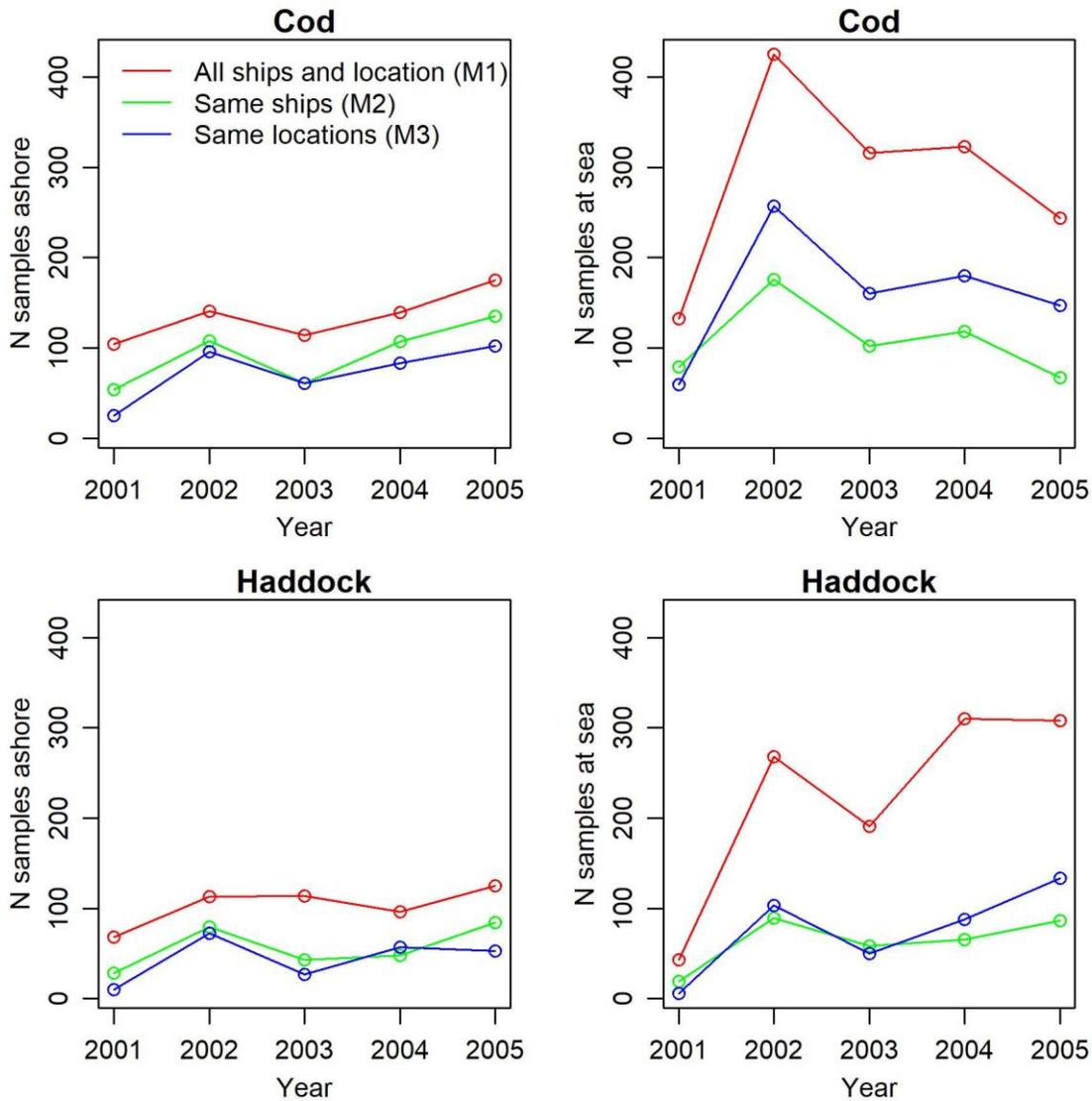

**Figure 4.** Number of samples at sea and ashore used for the estimation of discard rate for cod and haddock for the three methods: M1) All ships and locations, M2) same ships, and M3) same locations.

Three bootstrapping methods to estimate the confidence interval were compared for each data usage method (Fig. 5). Assuming that all observations (fish length measurements) where independent resulted in the narrowest confidence interval. This assumption is however not valid as it ignores the intra-class correlation and gives unrealistically low uncertainty in the discard estimate acquired from survey data and should not be used. There was not much difference between the other two methods, i.e. bootstrapping at all levels and bootstrapping at the highest level (ship). Bootstrapping at all levels gave a slightly wider confidence interval than bootstrapping at the highest level for both cod and haddock and all three data usage methods (Fig. 5).

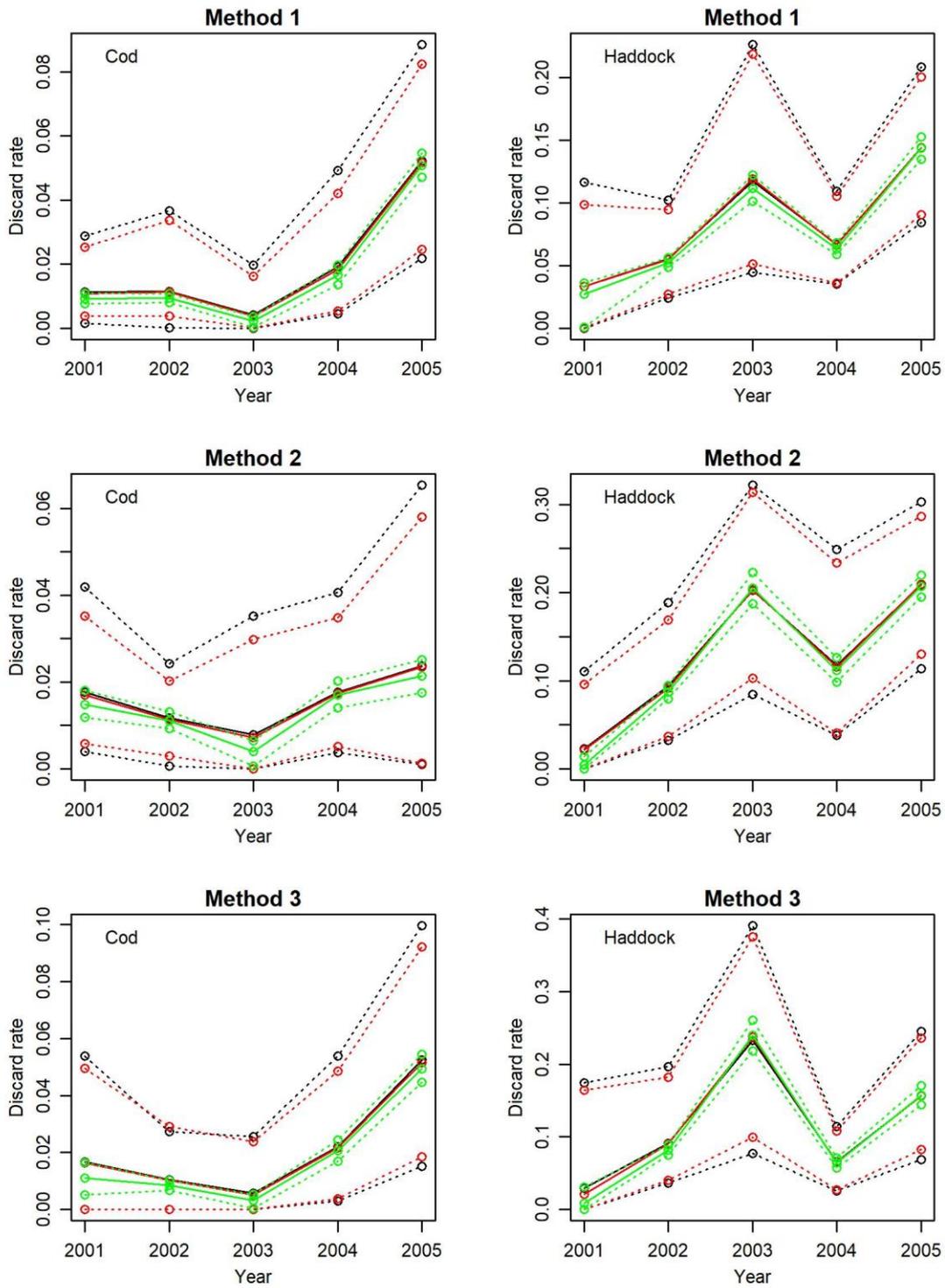

Figure 5. Comparison of the three bootstrapping methods: a) resampling at all three levels (black), b) resampling at the highest level (red), and c) resampling assuming independent observations (green). The comparison was done for the three data usage methods for the estimation of discard rate by numbers for both cod and haddock.

## 3.2 *The effect of sample sizes*

### 3.2.1 The precision of the estimate

A simulation was carried out to see the effect of changing the number of ships, number of samples taken from each ship and number of observations from each sample on the precision of the discard rate estimate. The results showed that the precision increases, i.e. the confidence interval becomes smaller, as the number of ships used in the data collection increases, the precision also improves with increasing number of samples (Fig. 6). The increase in number of samples is however more important when there are fewer ships. For example the range of the confidence interval of the cod discard rate decreased by 20% when the sample size goes from one to two samples for ten ships but the decrease is only 8% for 40 ships. The effect of increasing the number of observations from each sample is also more important when the number of ships are few, the range of the confidence interval of the cod discard rate decreased by 5% when the number of observations went from 25 to 100 for 10 ships but only by 1% when there were 40 ships.

The results from the simulations also showed that increasing the number of ships at sea improves the precision of the estimate more than increasing the number of samples taken from ships. When observations were sampled from ten ships increasing the sample size to two tows, having 20 samples, each with 25 fish, resulted in 22% decrease in the range of the confidence interval for the discard rate of cod but increasing the number of ships to 20, again having 20 samples each with 25 fish, led to 28% smaller confidence interval. This effect was the same for 20 ships, where increasing the number of samples to two led to 17% decrease in the range of the confidence interval but doubling the number of ships resulted in 21% smaller confidence interval. For the precision of the discard rate of haddock it seems that increasing the number of ships is much more beneficial than increasing the number of samples from each ship. Increasing the number of samples to two for 10 ships decreased the confidence interval by 8% but increasing the number of ships from 10 to 20 decreased the confidence interval by 21%. This is consistent with result from Helle and Pennington (2004) which showed that when estimating the mean length of fish the variance is dominated by the ship component.

It gave more precise estimate to increase the number of ships sampled at sea than ashore. The confidence interval decreased by 25% for cod discard rate and 23% for haddock when the number of ships at sea were increased from 10 to 20 but by 17% and 13% for the same increase in ships ashore for cod and haddock respectively. It gave more improvement increasing the samples taken from each ship at sea than ashore for the estimate of cod discard rate. Increasing the sample size to two at ships at sea decreased the confidence interval by 18% but only by 10% ashore but there was little difference in the case of the haddock where the increase in sample size led to 7% decrease for the samples at sea and 6% decrease for the samples taken ashore.

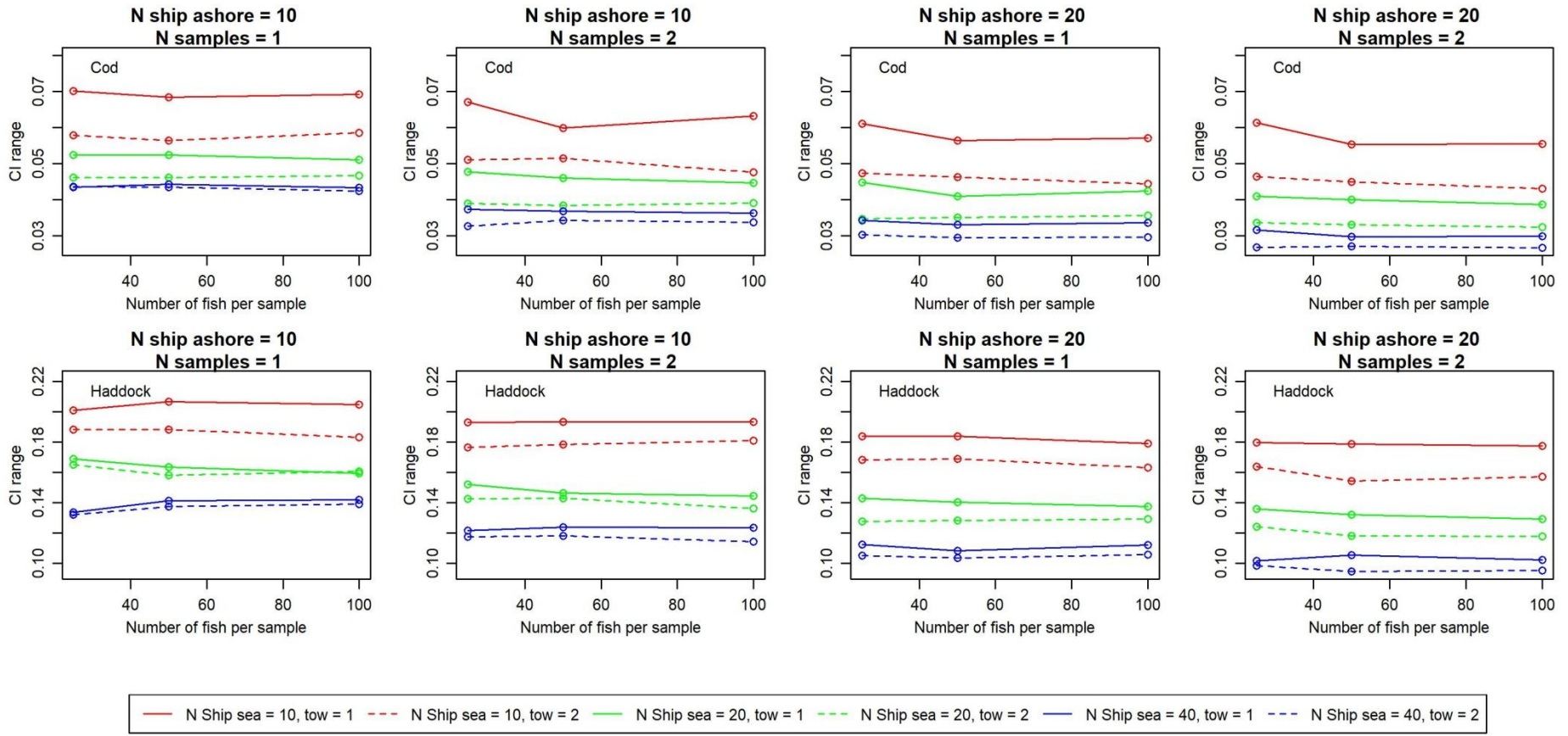

Figure 6. The range of the 95% confidence interval (CI) of the discard rate estimate from the Monte Carlo simulations using all available data from the year 2002.

## 3.2.2 The bias of the estimate

The estimated average discard rate was overestimated when the sampling size was small (Fig. 7). The bias of the stimates comes from from Eq. 5 where the denominator often becomes zero as no fish of that length group was sampled from the catch. The values for those length groups can then not be determined. This occurs more frequently for the small or large length groups as fish from those length groups is more rare in the catch that the intermediate lengths. For example in the case with the smallest sample size the estimated discard rate was 73% higher for cod and 22% for haddock than where the sample size was the largest. Increasing the number of observations from each sample reduced the bias for the cod. The cod has higher length range than the haddock which leads to more undetermined values than for the haddock which then results in higher bias for the cod.

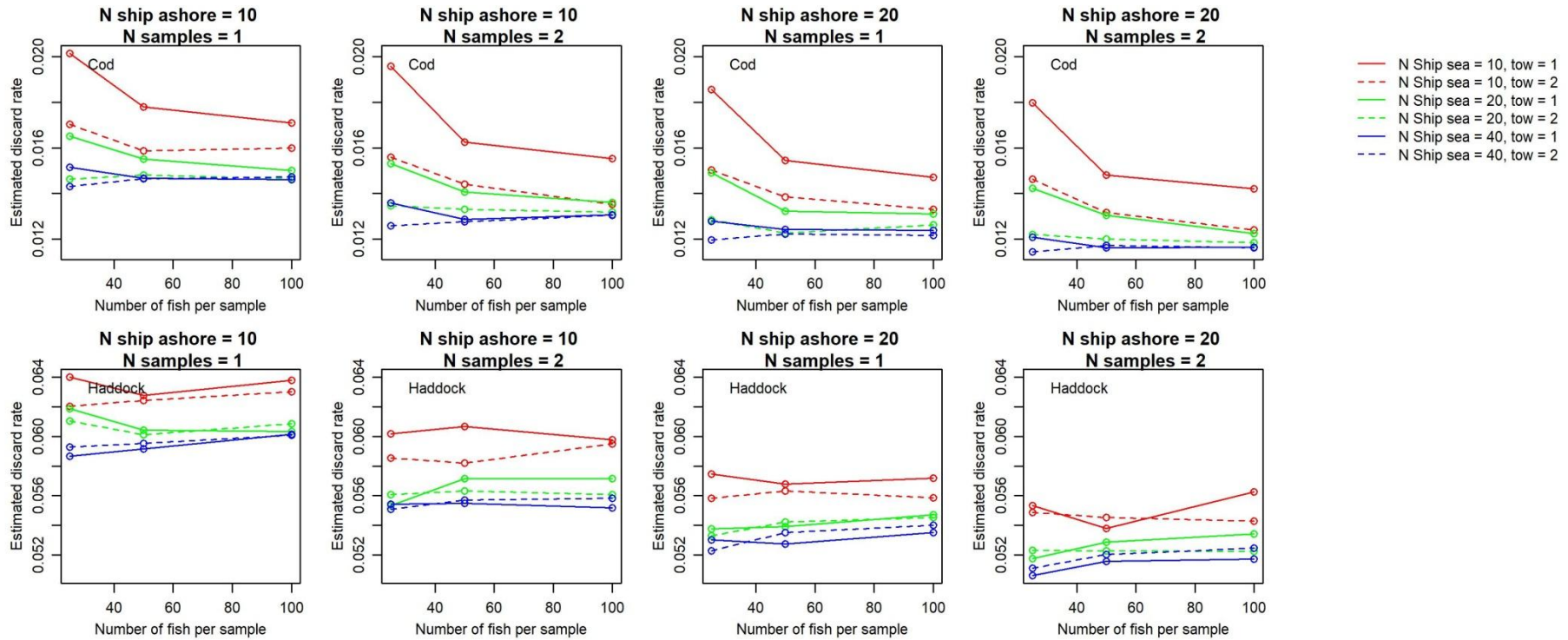

Figure 7. The average estimated discard rate of cod and haddock from the Monte Carlo simulations using all available data from the year 2002.

Instead of having undetermed values in Eq. 5 it is possible to put them to zero instead. It is then assumed that no fish is caught at this length and therefore also that no fish is discarded. This practise leads to reduced bias of the estimated discard rate acheived from small samples (Fig. 8). The standard method includes length groups of 1 cm. Increasing the range of each length groups also reduces the bias because it is then more likely that the length group includes observations from the catch, i.e. non-zero value in the denomiator.

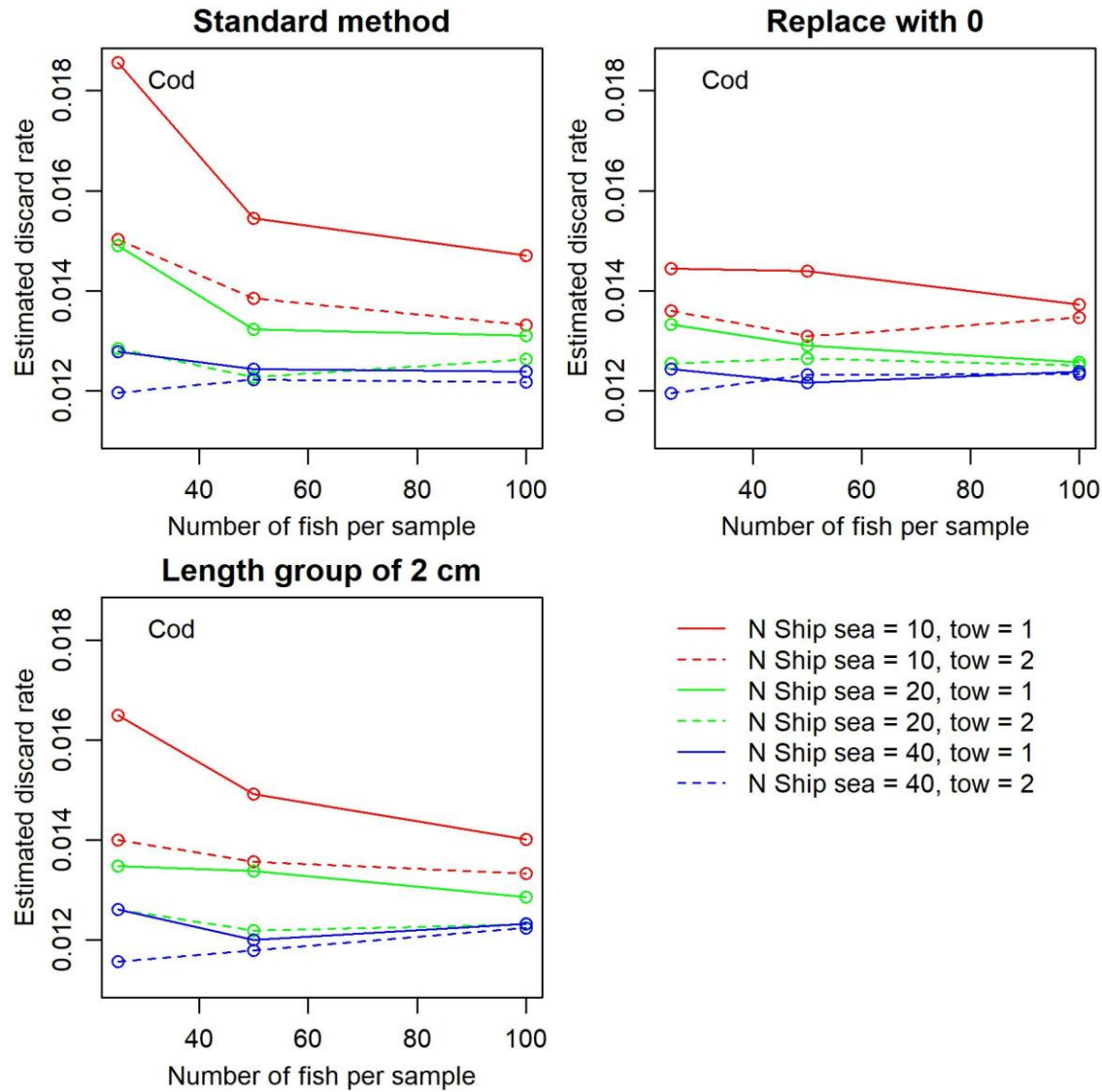

Figure 8. Bias of the discard rate estimate for: the standard method with undetermined values in Eq. 5 , where undetermined values are replaced with 0, and where length groups are of 2 cm for different samples sizes.

# Conclusion

It is possible to estimate a discard rate where a discard ban is in place. It however requires extensive monitoring in order to get an accurate estimate. It should also be noted that this estimation method determines the minimal discard rate as it assumes that a fish above a certain length is not discarded and does therefore exclude all above quota discarding. It also does not include discarding that is practiced in fisheries that target other species and may catch the discarded species as a by-catch. It is important to account for the intra-class correlation and the structure of the data when estimating the uncertainty of the estimated discard rate. This structures need to be considered when evaluating monitoring schemes. It can therefore be recommended to start with an extended monitoring scheme when setting up a monitoring program for discarding. A simulation study such as carried out in the present study can then be done to evaluate which sample sizes are adequate to estimate the discard rate with sufficient uncertainty.